        \documentstyle[preprint,tighten,eqsecnum,aps]{revtex}
        
\begin{document}
        \draft
	\preprint{UTS-DFT-95-4}

	\title{String Propagator: a Loop Space Representation}

	\author {S.Ansoldi\footnote{E-mail address:
	ansoldi@vstst0.ts.infn.it}}
	\address{ Dipartimento di Fisica Teorica
	dell'Universit\`a,\\
	Strada Costiera 11, 34014-Trieste, Italy}

	\author{ A.Aurilia\footnote{E-mail address:
	aaurilia@csupomona.edu}}
	\address{ Department of Physics, California State
	Polytechnic
	University\\
	Pomona, CA 91768}

	\author{ E.Spallucci\footnote{E-mail address:
	spallucci@vstst0.ts.infn.it}}
	\address{Dipartimento di Fisica Teorica dell'Universit\`a,\\
	Istituto Nazionale di Fisica Nucleare, Sezione di
	Trieste,\\
	Strada Costiera 11, 34014-Trieste, Italy}
	\maketitle

	\begin{abstract}
	The string quantum kernel is normally written as a
	functional sum over the string coordinates and  the world--sheet
	metrics. As an alternative to this quantum field--inspired
	approach, we study the
	closed bosonic string propagation amplitude in the functional
	space of loop configurations. This functional theory is based
	entirely on the Jacobi variational formulation of quantum
	mechanics, {\it without the use of a lattice approximation}.
	The corresponding
	Feynman path integral is weighed by a string action which is
	a {\it reparametrization invariant} version of the Schild
	action. We show that this path integral formulation is
	equivalent to a functional ``Schrodinger''
	equation defined in loop--space.
	Finally, for a free string, we show that the path integral
	and the functional
	wave equation are {\it exactly } solvable.
        \end{abstract}

		\section{Introduction}

	There are at least two approaches to the quantum theory of
	relativistic strings. One way is to look at a string model as a
	field theory in two spacetime dimensions. In this case, the string
	coordinates $x^\mu(\tau,\sigma)$ are reinterpreted as a
	multiplet of
	scalar fields defined over the string manifold parametrized by a
        Lorentzian coordinate mesh $(\tau,\sigma)$. The non--linearity of
	the Nambu--Goto action can be ``softened'' by assigning
	an auxiliary metric field $\gamma_{ab}(\tau,\sigma)$ over the string
	manifold, and then writing the action in the Howe-Tucker
	form \cite{ht}. After this reshuffling of variables,
	the original string model is converted into a
	{\it local field theory} and is quantized through canonical,
        or path integral methods \cite{book}.
	Quantum fluctuations around
        a classical solution eventually give rise to a spectrum of
	elementary particles,
        and the string itself acquires the status of fundamental
	building block of everything in the universe.

	On the other hand, one may regard a string as an elementary
	physical system  by itself, and focus on the geometric and
	topological properties of the string manifold.
	Vortices in a super--conducting medium \cite{abriko}
        and cosmic strings \cite{kibble} are two noteworthy
	examples of this geometrical approach. Quantum fluctuations
	are now interpreted as transitions between different string
	configurations. In particular, the quantum propagation kernel
	acquires the meaning
	of probability amplitude for the string shape to evolve from
	an initial configuration, represented by the non
	self--intersecting spatial loop
	$C_0: 0\le s\le 1 \rightarrow x^\mu_0=x^\mu(s), \,
	x^\mu_0(0)=x^\mu_0(1)$, to a final, non self--intersecting
	 configuration $C:
	x^\mu=x^\mu(s)$.\\ Thus, in this functional approach, spatial
	deformations of the string shape are mapped into ``translations''
	in the space of all possible loop configurations, and our major
	concern is to develop a ``Hamiltonian'' theory for the quantum
	mechanics of strings in loop space. As a matter of fact,
	the main purposes of this paper can be stated as follows: \\
	i) to give a path integral definition for the string kernel; \\
	ii) to prove that the path
        integral definition is equivalent to a functional wave
	equation defined over a ``space of loops'', and to show that the same
	kernel can be derived from both formulations; \\
	iii) to determine the exact form of the free string
	propagation kernel. \\
	Accordingly, the paper is organized as follows.\\
	In Sect.2, we obtain the Jacobi functional equation for a
	closed string.
	This equation is the essential link between the classical and
	quantum theory because it provides the WKB approximation to the
	whole functional ``Schrodinger'' equation.\\
	In Sect.3, we define the Feynman path integral for the quantum
	propagation kernel. We find that the
	reparametrization invariant sum over world-sheets, weighed by
	the exponential of the Nambu--Goto action, is a Jacobi path
	integral, i.e. a sum over string histories at ``fixed energy''
	$E=1/4\pi\alpha'$.\\
	In Sect.4, we present a path integral derivation of the string
        kernel wave equation. No discretization procedure is
	involved. Instead, we use Jacobi's variational principle
	to derive the functional Schrodinger equation for
	the string in a
	manifestly reparametrization invariant form.\\
	In Sect.5, we compute the quantum kernel for a free string in
	two different ways. In Subsct.5.1, we solve the kernel
	``Schrodinger''
	equation by exploiting the role of the Jacobi equation as the
	classical
        limit of the full quantum equation. In Subsct.5.2, the
	string quantum
	kernel is obtained from the path integral using a ``trick''
	which bypasses the use of semi--classical approximation or
	phase space discretization.\\
	Sect.6 is devoted to a brief summary of the results and to the
	discussion of their possible generalization.

	\section{ Functional Jacobi equation}
 	\label{class}
	As a starting point for the study of string dynamics one can
	choose either the Nambu--Goto action, or the Schild action: both
	functionals
        lead to the same classical dynamics \cite{schild}.
	It is not clear, however, whether or not such an equivalence
	persists unrestricted at
        the quantum level. This is because, in a quantum theory, the
	string propagation kernel reflects the different weight
	assigned to
	the string trajectories in the two classical frameworks.
	Another
	potential source of inequivalence stems from the fact that the
	Nambu--Goto action is reparametrization invariant but non linear
	with respect to the generalized velocity, whereas the Schild action
	is linear but at the expense of reparametrization invariance. Apart
	from all this, the standard procedure
	to construct the path integral in quantum mechanics applies to
	quadratic actions,
	which is not the case for relativistic systems.  One way to
        deal with the problem would be to follow the Dirac
	quantization procedure for constrained
        systems. But then, a canonical evolution of the system
        does not make sense because of the vanishing of the Hamiltonian.
	To preserve a Hamiltonian--type evolution, it is necessary to
	start with a non--reparametrization invariant theory. Even in this
	case, the resulting dynamics for an extended
	object is non--canonical.\\
	All of the above arguments converge to the focal question: if one
	insists on a Hamiltonian, albeit non canonical formulation of string
	dynamics, is there an evolution parameter which plays the role of
	``time variable'', and is this choice consistent with
	reparametrization invariance?\\
	Previous attempts to deal with those questions lead to seemingly
	conflicting conclusions. For instance, the string propagator
	obtained in ref.\cite{egu}, has been criticized in ref.\cite{ogie}.
	From our vantage point, the critical issue is that of
	reparametrization invariance. Both authors are led to a string
	diffusion equation which is manifestly {\it dependent} on the string
	parameter $s$, leaving us with the impression that the lack of
	reparametrization invariance
        of the classical action manifests itself even at the quantum
	level. However, in ref.\cite{egu},
	the physical Green function is obtained
	by averaging over all the possible values of the proper
	evolution parameter. In ref.\cite{ogie}, instead, it is
	claimed that the parametric dependence of the propagation
	kernel is  only apparent
	because the action is insensitive to the location of the area
	increment along the world--sheet boundary.
	The resulting wave equation is {\it local}, in the sense that
	it is defined
	at a single, representative, point on the string loop, and
	does not apply to the string as a whole. Evidently, in this
	approach
	the string is treated as a collection of constituent points, and
	this may well be a viable interpretation. However, inspired by our
	previous work on the classical dynamics of p--branes\cite{noi1},
	\cite{noi2},\cite{noi3}, we believe that the
	dynamics of each individual point on the string does not give a
	consistent account of the dynamics of the whole string.\\
	The alternative point of view is that ``the whole string is
	more than the sum of its parts'', and in this paper we wish to
	suggest a different approach
	which, in our view, addresses directly the question of the
	choice of dynamical variables and the related issue of
	reparametrization invariance in the classical theory as well as in
	the quantum theory. The stipulation is also made that the
	classical
	theory must emerge as a well defined limiting case of the
	quantum
	theory. In order to fulfil this condition, we invoke a {\it single
	dynamical principle} encompassing both areas of string--dynamics,
	namely the Jacobi variational principle suitably adapted to the case
	in which the physical system is a relativistic extended object.
	Thus, the dynamical variables are restricted to vary within the
	family of string trajectories which are solutions of the classical
	equation of motion. In other words, the variational procedure
	applies only to the final configuration of the string, rather than
	to its spacetime history.

	Against this conceptual backdrop, the {\it formalism} developed in
	this paper, largely inspired by the work of Nambu
	\cite{nambu1},\cite{nambu2} and Migdal \cite{migdal}, fully reflects
	our emphasis on the global structure of the string: our action
	functional is a {\it reparametrized} form of the
        Schild action, manifestly invariant under general coordinate
	transformation in the string parameter space, while preserving the
	polynomial structure in the dynamical
	variables; the natural candidate for the role of {\it time
	variable} is the proper area of the string world--sheet
	(equation 2.8), i.e., the invariant measure of the model manifold
	representing
	the evolution of the string. The final outcome is
	a {\it manifestly reparametrization
	invariant Schrodinger equation} which has the same form of
	the corresponding equation obtained from the Nambu--Goto action
	using a lattice approximation, and admits gaussian type
	wave packets as solutions.\\
	Our starting point is the Schild string action in Hamiltonian
	form
	\begin{eqnarray}
	S[x(\xi),p(\xi)]&=&
	{1\over 2}\int_{X(\xi)}p_{\mu\nu}dx^\mu\wedge dx^\nu-
	\int_\Sigma d^2\xi\,H(p)\\
	H(p)&\equiv& {1\over 4m^2}p_{\mu\nu}p^{\mu\nu}\ ,
	\label{act}
	\end{eqnarray}
	where $m^2=1/2\pi\alpha'$ is the string tension, $\Sigma$
	represents the model manifold of the string in parameter space, and
	$X(\xi)$ represents its image in Minkowski space. Then,
	$p_{\mu\nu}$
        stands for the linear momentum canonically conjugated to
        the world--sheet tangent element $\dot x^{\mu\nu}\equiv
	\epsilon^{ab}\partial_a x^\mu\partial_b x^\nu$.
	Both variables were
	originally introduced by Nambu \cite{nambu1},
	\cite{nambu2}. More recently, the same variables were used to
	formulate a gauge theory for the
        dynamics of strings and higher dimensional extended objects
	\cite{noi1}, \cite{noi2}, \cite{noi3}. \\
	In order to cast the action (\ref{act}) in a reparametrization
        invariant form, we introduce a new pair of world--sheet
        coordinates $(\sigma^0,\,\sigma^1)$ through the
	boundary preserving
	transformation $\xi^a\rightarrow \sigma^a=\sigma^a(\xi)$,
	and promote the original pair $(\xi^0,\,\xi^1)$
	to the role of dynamical variables.
	Then, $S[x(\xi),p(\xi)]$ transforms into
 	\begin{equation}
	S[x(\sigma),p(\sigma), \xi(\sigma)]=
	{1\over 2}\int_{X(\sigma)}p_{\mu\nu}dx^\mu\wedge dx^\nu-
	  {1\over 2}\epsilon_{ab}\int_{\Sigma(\sigma)}
	d\xi^a \wedge d\xi^b\, H(p)
	\label{actr}
	\end{equation}
	The new action (\ref{actr})
	 is numerically equivalent to (\ref{act}) and leads to
	the same equation of motion for $x^\mu$, $p_{\mu\nu}$.
	Furthermore,
	 variation with respect to the new fields $\xi^a(\sigma)$
	leads to the energy--balance equation
        \begin{equation}
	\epsilon_{ab}\epsilon^{mn}\partial_m\xi^a\partial_n H=0 \ ,
	\Longrightarrow
	H_{\rm cl.}={\rm const.}\equiv E
	\label{balance}
	\end{equation}
	which, in our case, correctly shows that the Hamiltonian is
	constant along a classical solution.
	The action (\ref{actr}) is linear with respect to the ``velocities''
        $\dot x^{\mu\nu}$ and $\dot\xi^{ab}(\equiv
	\epsilon^{mn}\partial_m \xi^a\partial_n\xi^b)$.
	Hence, if one interprets
	$\pi_{ab}\equiv \epsilon_{ab}H$
	as the momentum canonically conjugated to $\xi^a(\sigma)$,
	then (\ref{actr}) acquires the form of a reparametrization invariant
        theory in six dimensions \cite{nambu1}.\\
	The Jacobi equation for the string is obtained
	by varying $S[x(\sigma),p(\sigma), \xi(\sigma)]$
	within the family of world--sheets which solve the string
	equations of motion.
	We emphasize that this type of variation corresponds to a
	deformation of the only
	free boundary of the world--sheet, i.e. $C$, and corresponds
	to the more familiar variation of the world--line end--point in the
	case of a particle. Then, the steps leading to the Jacobi equation
	are as follows. First, the contribution
        from the variation of the world--sheet itself vanishes by
	definition, and we obtain:
	\begin{equation}
	\delta S_{\rm cl.}[\partial X; A]=\oint_{\partial X}p_\mu\,
	\delta x^\mu -H_{\rm cl.} \delta A  .
	\label{hjvar}
	\end{equation}
	Next, we note that in view of the constancy of the Hamiltonian
	over a classical trajectory, we can vary the area of the $\Sigma$
	domain without reference to
        the specific point along the boundary $\partial\Sigma$ where the
	infinitesimal variation takes place. In other words, we can
	move ``$\delta$''in front of the area integral and then trade the
	functional variation $\delta A$ for an ordinary differential
	variation $dA$, and define
	\begin{equation}
	p_\mu(s)\equiv {\delta S_{\rm cl.}\over \delta x^\mu(s)}=
	p_{\mu\nu}\, x^{\,\prime\,\nu}
	\label{bmom}
	\end{equation}
	as the boundary momentum density. Similarly, the area--energy
	density $E$ can
	be written as the partial derivative of the classical action
	with respect to the invariant measure of the $\Sigma$ domain
	in parameter space:
	\begin{eqnarray}
	&&E=-{\partial S_{\rm cl.}\over \partial A}\\
	&&A\equiv {1\over 2}\epsilon_{ab}\int_\Sigma d\xi^a\wedge
	d\xi^b\ .
	\end{eqnarray}
	 Hence, the Jacobi variational principle in the form of
	equation (\ref{hjvar})
	shows that $ p_\mu(s)$ is conjugated to the spacetime
	world--sheet boundary variation, while $H_{\rm cl.}$ describes the
	response of the classical
        action to an arbitrary area variation in parameter
	space. Thus, if we consider string dynamics from the loop space
	point of view \cite{migdal}, then $A$ and $x^\mu(s)$ can be
	interpreted as the
	``{\it time}'' and ``{\it space}'' positions of the final
	string $C$
	with respect to the initial one $C_0$, which we assume to be
	fixed at the outset. In this perspective,
	$H_{\rm cl.}$ is the {\it area--hamiltonian, or generator} of the
	classical
	evolution from the ``initial time'' $T=0$ to the final time $T=A$.
	Accordingly, $p_\mu(s)$ is the generator of
	infinitesimal ``{\it translations}'' in loop space, which are
	perceived as infinitesimal deformations $x^\mu(s)\rightarrow
	x^\mu(s)+\delta x^\mu(s)$ of the string shape in Minkowski space. \\
	Finally, note that in this formulation, $E$ represents the {\it
	energy per unit area} associated with an extremal world--sheet
         of the action (\ref{actr}), while $p_\mu(s)$ is the
	{\it momentum per unit length} of the string loop $C$.
	Therefore, the energy--momentum dispersion relation can be
	written either as an equation between {\it densities}
	\begin{equation}
	{1\over 2m^2}p_\mu p^\mu={1\over 4m^2} p_{\mu\nu} p^{\mu\nu}
	x^{\,\prime\, 2}(s)=E  x^{\,\prime\, 2}(s)
	\label{constr}
	\end{equation}
	or, as an {\it integrated} relation
	\begin{equation}
	{1\over 2m^2}\int_0^1{ds\over\sqrt{x^{\,\prime\,2}}}\,
	p_\mu p^\mu=E \int_0^1 ds\sqrt{x^{\,\prime\,2}}\ .
	\label{dispersion}
	\end{equation}
	The above equation, once written in terms of $S_{\rm cl.}$, turns into
	the promised functional Jacobi equation for the string:
	\begin{equation}
	\left(\int_0^1ds\,\sqrt{x^{\,\prime\,2}} \right)^{-1}
	\int_0^1{ds\over\sqrt{x^{\,\prime\,2}}}\,
	{\delta S_{\rm cl.}\over \delta x^\mu(s) }{\delta S_{\rm cl.}\over
	\delta x_\mu(s) }=
	-2m^2{\partial S_{\rm cl.}\over\partial A}\ .
	\label{hjf}
	\end{equation}
	Looking in more detail at this equation, we observe that the
	covariant integration over $s$
	takes into account all the possible locations
	of the  point, along the contour $C$,
	where the variation can be applied. But, in this
	 way, every point of $C$ is overcounted a ``number of times''
        equal to the string proper length. The first factor, in
	round parenthesis, is just the string proper length and removes such
	overcounting. In other words, we sum over all the possible
	ways in which one can deform the string loop, and then divide by the
	total number of them. The net result is that
	the l.h.s. of equation(\ref{hjf}) is insensitive
        to the choice of the point where the final string $C$ is
        deformed. Therefore the r.h.s. is a genuine
	reparametrization
   	scalar which describes the system's response to the extent of
	area variation, irrespective of the way in which the
	deformation is implemented.
	With hindsight, the wave equation proposed in
	\cite{egu},\cite{ogie} appears
        to be more restrictive than equation (\ref{hjf}), in the
	sense that it
	requires the second variation of the line functional
	to be proportional to $x^{\prime\,2}(s)$ at {\it any} point on
	the string loop, in contrast to equation (\ref{hjf}) which represents
	an integrated constraint on the string as a whole.\\
	Equation (\ref{hjf}) is the starting point in the first
	quantization program via the Correspondence Principle:
        one introduces the quantum operators
	\begin{equation}
	\widehat p_\mu(s)\equiv i\hbar {1\over\sqrt{x^{\,\prime\,2}} }
	{\delta\over\delta x^\mu(s)}\ ,\quad
	\widehat H\equiv -i\hbar {\partial\over\partial A}\ ,
	\end{equation}
	and imposes the operatorial form of the
	dispersion relation (\ref{dispersion}) on the string
        wave functional $\psi[C; A]$.
	Alternatively, one can focus directly on
	the string propagation kernel $K\left[x(s),x_0(s); A\right]$,
	 in which case we turn to Feynman's
        ``sum over histories'' method since this is probably the
	most natural and
        effective way to define $K\left[x(s),x_0(s); A\right]$ in
	quantum string--dynamics.

	\section{Feynman and Jacobi path integrals}

	The path integral approach to string--dynamics is not only a
	useful check of the
	quantization procedure, but gives a new insight into
        the quantum theory itself. In particular, it gives a better
	insight into the
	meaning of the equivalence between the Nambu--Goto action and the
	Schild action. Furthermore,
	it provides a physically transparent relationship between the
	Feynman ``sum over histories'' integral and the Jacobi path
	integral. However, to see how this comes about, one must keep in
	mind  that the momenta $p_{\mu\nu}$
        and $\pi_{ab}$ cannot vary freely, because $\pi_{ab}$ is
	merely a shorthand notation for the function $\epsilon_{ab}H(p)$,
	and thus must satisfy the constraint equation
	\begin{equation}
	\pi_{ab}-\epsilon_{ab}H(p)=0\ .
	\label{vincolo}
	\end{equation}
	This constraint may be incorporated in the action (\ref{actr})
	by means of a Lagrange multiplier $N^{ab}(\sigma)$
	\begin{eqnarray}
	&&S[x(\sigma),p(\sigma), \xi(\sigma), N(\sigma); A]=
	{1\over 2}\int_{X(\sigma)}p_{\mu\nu}dx^\mu\wedge
	dx^\nu+\nonumber\\
	&&{1\over 2}\int_{\xi(\sigma)}\pi_{ab}d\xi^a\wedge
	d\xi^b-
	{1\over 2}\int_\Sigma d^2\sigma N^{ab}(\sigma)
	\left[\pi_{ab}-\epsilon_{ab}H(p)\right]\ ,
	\end{eqnarray}
	and its physical meaning can be read out of the classical
	equation of motion obtained by varying $\pi_{ab}$:
	\begin{equation}
	N^{ab}(\sigma)=\epsilon^{mn}\partial_m \xi^a\partial_n
	\xi^b\ .
	\end{equation}
	Thus, $ N^{ab}$ is the transformed $\epsilon$--tensor in the
	new coordinate system. Then, apart from an over all
	normalization constant, \footnote{The normalization
	constant will be fixed after all the functional integrations
	are carried out, by imposing the boundary condition
	\begin{equation} lim_{A\rightarrow
	0}K[x(s),x_0(s);A]=\delta[x(s),x_0(s)]\ .
	\label{bcond}
	\end{equation}
	Such a procedure effectively amounts to a renormalization
	of the field--dependent determinants produced by gaussian
	integration. }the amplitude
        for the initial string $C_0$ to ``evolve'' into the final
        string $C$ in a lapse of ``time $A$'', can be represented by the
         path integral
	\begin{eqnarray}
	&&K[x(s),x_0(s);A]=\int_{x_0(s)}^{x(s)}
	\int_{\xi_0(s)}^{\xi(s)}[D\mu(\sigma)]
	e^{iS[x,p,\xi,\pi,N;A]/\hbar}
	\nonumber\\
	&&[D\mu(\sigma)]\equiv [Dx^\lambda(\sigma)][D\xi^a(\sigma)]
	[Dp_{\mu\nu}(\sigma)][D\pi_{ab}(\sigma)][DN_{cd}(\sigma)] .
	\label{pathint}
	\end{eqnarray}

	At the classical level, $H$, or $\pi_{ab}$, is independent of
	the coordinates $\sigma^m$  because of the balance equation
	(\ref{balance}). The
        same result is obtained at the quantum level by integrating
	out the $\xi^a(\sigma)$ fields:
	\begin{eqnarray}
	&&\int_{\xi_0(s)}^{\xi(s)}[D\xi^a(\sigma)]
	\exp\left\{{i\over 2\hbar}
	\int_{\xi(\sigma)}d\xi^a\wedge
	d\xi^b\,\pi_{ab}\right\}=\nonumber\\
	&&\delta\left[\epsilon^{mn}\partial_m\pi_{an}\right]\exp
	{i\over 2\hbar}\int_{\xi(\sigma)}d\left(\pi_{ab}\xi^a
	d\xi^b\right) .
	\label{nonumber}\\
	\end{eqnarray}
	The functional Dirac--delta requires $\pi_{ab}$ to satisfy the
	classical equation of motion, i.e.
	$\pi_{ab}=\epsilon_{ab}\times
	{\rm const.}\equiv \epsilon_{ab}E$. Therefore,
	\begin{eqnarray}
	&&\int
	D[\pi_{ab}]\delta\left[\epsilon^{mn}\partial_m\pi_{an}
	\right]
	\exp \left\{{i\over 2\hbar}\left[\pi_{ab}\int_{\xi(\sigma)}
	d\left(\xi^a d\xi^b\right)-
	\int_\Sigma d^2\sigma
	N^{ab}(\sigma)\pi_{ab}\right]\right\}=\nonumber\\
	&&\int_0^\infty dE e^{iEA/\hbar}\exp \left\{-{iE\over 2\hbar}
	\int_\Sigma d^2\sigma N^{ab}(\sigma)\epsilon_{ab}\right\}
	\end{eqnarray}
	where we have assumed that the Hamiltonian is bounded from
	below and is normalized in such a way that $E\ge 0$.
	Then, the Feynman path integral can be written as follows

	\begin{eqnarray}
	&&K[x(s),x_0(s);A]=\int_0^\infty dE e^{iEA/\hbar}
	\int_{x_0(s)}^{x(s)}[Dx^\mu(\sigma)][Dp_{\mu\nu}(\sigma)]
	[DN_{cd}(\sigma)]\times\nonumber\\
	&&\exp\left\{
	{i\over 2\hbar}\int_{X(\sigma)}p_{\mu\nu}dx^\mu\wedge dx^\nu
	-{i\over 2\hbar}\epsilon_{ab}\int_\Sigma d^2\sigma
	N^{ab}(\sigma)
	\left[E-H(p)\right]\right\}\nonumber\\
	&&\equiv 2i\hbar m^2\int_0^\infty dE
	e^{iEA/\hbar}G[C,C_0;E]\ ,
	\label{amplitude}
	\end{eqnarray}
	where we have introduced the {\it Jacobi path integral}
	$G[C,C_0;E]$
	as the {\it amplitude for a string to propagate from $C_0$ to
	$C$, at fixed energy $E$.}

	The most important property of
	$G[C,C_0;E]$ is {\it reparametrization invariance}.
	If we integrate out the area momentum $p_{\mu\nu}$, we obtain
	\begin{eqnarray}
	G[C,C_0;E]&&=\int_{x_0(s)}^{x(s)}[Dx^\mu(\sigma)][DN(\sigma)]
	\times\nonumber\\
	&&\exp\left\{ -{i\over\hbar}\int_\Sigma d^2\sigma\left[
	-{m^2\over 4N}{\dot x}^{\mu\nu}{\dot x}_{\mu\nu}+NE\right]\right\}
	\label{kjinv}
	\end{eqnarray}
	where $N(\sigma)\equiv \epsilon_{ab}N^{ab}(\sigma)/2$.
	Equation (\ref{kjinv})
	is manifestly invariant under reparametrization
	\begin{equation}
	{\dot x}^{\mu\nu}(\sigma)\rightarrow
	{\rm det}\left({\partial u^a\over\partial \sigma^m}\right)
	{\dot x}^{\mu\nu}(u)\ ,\quad N(u)\rightarrow
	 {\rm det}\left({\partial u^a\over\partial \sigma^m}\right)
	N(\sigma)\ .
	\end{equation}
	Finally, if we estimate the path integral (\ref{kjinv}) around the
	saddle point\\
	$\widehat N(\sigma)=\left(-m^2{\dot x}^{\mu\nu}{\dot x}_{\mu\nu}/4E
	\right)^{1/2}$,
	we find
	\begin{equation}
	G[C,C_0;E]=\int_{x_0(s)}^{x(s)}[Dx^\mu(\sigma)]
	\exp\left\{-{i\over \hbar}\sqrt{m^2 E}\int_\Sigma
	d^2\sigma\sqrt{-{\dot x}^{\mu\nu}{\dot x}_{\mu\nu}}\right\}\ ,
	\label{kng}
	\end{equation}
	which is the usual path integral weighed by the
	Nambu--Goto action, once we fix $E=m^2/2$.\\
	The result (\ref{kng}) suggests the following concluding
	remarks for this section:\\
	i) equation (\ref{kng}) could be assumed at the outset and
	taken as a starting
        point for string quantization by means of functional
	techniques. The advantage of our derivation is that it clarifies the
	physical meaning of such a
	reparametrization invariant path integral:
	it represents the string propagation amplitude at {\it fixed
	``area--energy'' $E=1/4\pi\alpha'$.}\\
	ii) We can invert the Fourier transform (\ref{amplitude})
	and {\it define}
	the reparametrization invariant path integral in terms of the
	Feynman propagation amplitude at fixed ``area--lapse'' $A$
	\begin{equation}
	G[C,C_0;m^2]\equiv {1\over 2i\hbar m^2}
	\int_0^\infty dA e^{-im^2 A/2\hbar}K[x(s),x_0(s);A]\ .
	\label{laplace}
	\end{equation}
	Then, reparametrization offers  an alternative definition of the
	sum over histories: first, sum over all world--sheets of fixed area;
	then, integrate over all possible values of the world--sheet area.
	\\
	iii) Equation(\ref{amplitude}) represents the quantum
	counterpart of the
        classical equivalence \cite{schild} between the Nambu--Goto action
	and the Schild action.

	\section{The string kernel wave equation }

	The purpose of this section is to show how to derive
	the functional wave equation for $K[x(s),x_0(s);A]$ from the
	corresponding path integral.
	This equation describes how the string responds to a
	variation of the final boundary $\xi^a=\xi^a(s)$,
	just as the
	ordinary Schrodinger equation describes a particle reaction
        to a shift of the time interval end--point. As we have seen
	in section (\ref{class}),
	the string ``natural'' evolution parameter is the area $A$ of
        the string manifold, so that functional or area derivatives
	generate
        ``translations'' in loop space, or string deformations in
	Minkowski space. Thus, we expect the functional wave
	equation to be of order one in $\partial/\partial A$, and of
        order two in $\delta/\delta x^\mu(s)$, or $\delta/\delta
	\sigma^{\mu\nu}(s)$\footnote{Functional and area derivatives
	are related by \cite{migdal}
	\begin{equation}
	{\delta\over\delta x^\mu(s)}=x^{\,\prime\,\nu}{\delta\over\delta
	\sigma^{\mu\nu}[C]}\ ,\quad x^{\,\prime\,\nu}\equiv
	{dx^\nu\over ds}
	\nonumber\ .
	\end{equation}
	Therefore, the functional wave equation can be written in
	terms of either type of derivative.
	Contrary to the statement in ref. \cite{ogie}, area derivatives
	are regular even when ordinary functional derivatives are not
	\cite{migdal}.}.\\
	The standard procedure to arrive at the kernel wave equation
	goes through a recurrence relation satisfied by the
	discretized version of the Jacobi path integral
	\cite{kawai}, \cite{suga}. However, such a construction is well
	defined only when the action is a polynomial in the dynamical
	variables. For a non--linear action such
	as the Nambu--Goto area functional, a lattice definition of the path
	integral is much less obvious. Moreover,
	the continuum functional wave equation is
	recovered through the highly non--trivial limit of vanishing
	lattice step \cite{kawai}, \cite{suga}. In any case,
	the whole procedure seems disconnected from the classical
	approach to string dynamics, whereas we would like to see a logical
	continuity between quantum and classical dynamics. Against this
	background, it seems useful to offer an {\it alternative}
	 path integral derivation of the string functional wave equation
	 which is deeply rooted in the
	Hamiltonian formulation of string dynamics discussed in
	Sect.(\ref{class}), and is basically derived from the same Jacobi
	variational principle which we have consistently adopted so far. \\
	The kernel variation under infinitesimal deformations of the
	field variables is
	\begin{equation}
	\delta K[x(s),x_0(s); A]={i\over\hbar}
	\int_{x_0(s)}^{x(s)}
	\int_{\xi_0(s)}^{\xi(s)}[D\mu(\sigma)]\,
	\delta S\exp\left({iS/\hbar}\right)\ .
	\label{deltak}
	\end{equation}
	As usual, only boundary variations will contribute to
	equation (\ref{deltak}) if we restrict the fields to vary
	within the family
	of classical solutions corresponding to a given initial string
	configuration. Then,
	\begin{equation}
	\delta S_{\rm cl.}[C;A]=\oint_C p_{\mu\nu}\,
	\delta x^\mu(s)\, dx^\nu-E\, dA\ .
	\label{varscl}
	\end{equation}

	From equations (\ref{deltak}) and (\ref{varscl}), we obtain
	\begin{eqnarray}
	&&{\partial\over \partial A}K[x(s),x_0(s); A]=-{iE\over\hbar}
	K[x(s),x_0(s); A]\ ,\label{darea}\\
	&&{\delta\over \delta x^\mu(s)}K[x(s),x_0(s); A]=
	{i\over\hbar}\int_{x_0(s)}^{x(s)}
	\int_{\xi_0(s)}^{\xi(s)}[D\mu(\sigma)]
	p_{\mu\nu}x^{\,\prime\,\nu}
	\exp\left({iS/\hbar}\right)\label{functd}\ .
		\end{eqnarray}
	Then, by comparison of (\ref{darea}), (\ref{functd}) and
	(\ref{dispersion}), one obtains immediately the kernel wave
	equation
	\begin{eqnarray}
	-{\hbar^2\over 2m^2}\left(\int_0^1 ds
	\sqrt{x^{\,\prime\, 2}}\right)^{-1}
	\int_0^1 {ds\over\sqrt{x^{\,\prime\, 2}}}
	{\delta^2\over \delta x^\mu(s)\ \delta x_\mu(s)}
	K[x(s),x_0(s);A]&&
	=\nonumber\\
	i\hbar{\partial\over \partial A}K[x(s),x_0(s);  A]&&
	\label{waveq}
	\end{eqnarray}

	Thus, $K[x(s),x_0(s); A]$ can be determined either by solving
	the functional wave equation (\ref{waveq}), or by evaluating the path
	integral (\ref{pathint}).\\
	Once equation (\ref{waveq}) is given, it is straightforward to
	show that
	$G[C,C_0;m^2]$ satisfies the following equation
	\begin{equation}
	\left[- \hbar^2\left(\int_0^1 ds\sqrt{x^{\,\prime\, 2}}
	\right)^{-1}
	\int_0^1 {ds\over\sqrt{x^{\,\prime\, 2}}}
	{\delta^2\over \delta x^\mu(s)\ \delta x_\mu(s)} +m^4\right]
	G[C,C_0;m^2]=-\delta[C-C_0] .
	\label{green}
	\end{equation}
	Therefore, $G[C,C_0;m^2]$
	can be identified with the Green function for the string.

	\section{Computing the kernel}
	\subsection{Integrating the functional wave equation}

	It is possible to compute $K[x(s),x_0(s); A]$ exactly
	in the ``free'' case because the
	Lagrangian corresponding to the Hamiltonian (\ref{act}) is
	quadratic with respect to the generalized velocities
	${\dot x}^{\mu\nu}$.
	Previous experience with this class of Lagrangians
        suggests the following ansatz for the string quantum kernel:
	\begin{equation}
	K_0[x(s),x_0(s); A]=
	{\cal N} A^\alpha\exp\left(i I[x(s),x_0(s);A]/\hbar\right)\ ,
	\label{ansatz}
	\end{equation}
	where ${\cal N}$ is a normalization constant, and $\alpha$
	a real number.
	Substituting this ansatz into eq.(\ref{waveq}) gives two
        independent equations for the amplitude and the phase
	respectively,
	\begin{eqnarray}
	{2\alpha m^2\over A}&=&
	-\left(\int_0^1 ds\sqrt{x^{\,\prime\, 2}}\right)^{-1}
	\int_0^1 {ds\over\sqrt{x^{\,\prime\, 2}}}
	{\delta^2 I\over \delta x^\mu(s)\ \delta x_\mu(s)}\label{ampl}\ ,\\
	2m^2{\partial I\over \partial A}&=&
	 -\left(\int_0^1 ds\sqrt{x^{\,\prime\, 2}}\right)^{-1}
        \int_0^1 {ds\over\sqrt{x^{\,\prime\, 2}}}
        {\delta I\over \delta x^\mu(s)}{\delta I\over \delta x_\mu(s)}\ .
	\label{phase}
	\end{eqnarray}
	Comparing equations (\ref{phase}) and (\ref{hjf}), we see that
	$I=S_{\rm cl.}[x(s),\,x_0(s);A]$ and (\ref{phase}) is just the
	classical Jacobi equation.
	Therefore, the main problem is to determine the form of
        $S_{\rm cl.}$ in the string case. We do so by analogy with the
	relativistic point--particle case, where $S_{\rm cl.}$ is a
	functional of the world--line length element. Accordingly, we first
	introduce the {\it oriented
        surface element} as a functional of the surface boundary $C$
	\begin{equation}
	\sigma^{\mu\nu}[C]\equiv \oint_C x^\mu dx^\nu=\int_0^1 du\, x^\mu(u)
	{dx^\nu\over du}\ .
	\label{areael}
	\end{equation}
	Then, from the above definition we obtain
	\begin{eqnarray}
	&&{\delta \sigma^{\mu\nu}[C]\over \delta x^\alpha(s)}=
	\delta_\alpha{}^\mu x^{\,\prime\,\nu}(s)-
	\delta_\alpha{}^\nu x^{\,\prime\,\mu}(s)\ ,\label{f1}\\
	&&{\delta^2 \sigma^{\mu\nu}[C]\over
	\delta x^\alpha(s)\delta x^\beta(u)}=
	\left(\delta_\alpha{}^\mu
	\delta_\beta{}^\nu-\delta_\alpha{}^\nu\delta_\beta{}^\mu\right)
	{d\over ds}\delta(s-u)\ .\label{f2}
	\end{eqnarray}
	Next, we introduce the trial
        solution
	\begin{eqnarray}
	S_{\rm cl.}[x(s),x_0(s);A]&=&{\beta\over 4A}\left(\sigma^{\mu\nu}[C]-
	\sigma^{\mu\nu}[C_0]\right)\left(\sigma_{\mu\nu}[C]-
	\sigma_{\mu\nu}[C_0]\right)\ ,\nonumber\\
	&\equiv&{\beta\over 4A}\Sigma^{\mu\nu}[C-C_0]\Sigma_{\mu\nu}[C-C_0]
	\label{trial}
	\end{eqnarray}
	where $\beta$ is a second parameter to be fixed by the equations
	(\ref{ampl}), (\ref{phase}). By taking into account (\ref{f1}),
	(\ref{f2}), we find
	\begin{equation}
	{\delta S_{\rm cl.}\over\delta x^\mu(s)}=
	{\beta\over A}\Sigma_{\mu\nu}[C-C_0]\,x^{\prime\,\nu}(s)\ .
	\end{equation}
	Note that the dependence on the parameter $s$ is only through the
	factor $x^{\prime\,\nu}(s)$. Then,
		\begin{equation}
	{\delta^2S_{\rm cl.}\over \delta x_\mu(s)\delta x^\mu(s)}=
	{3\beta\over A}
	x^{\,\prime\,2}(s)\ .
	\end{equation}
	Equations (\ref{phase})and (\ref{ampl}) now give
	\begin{equation}
	\beta=-{2\alpha\over 3}m^2\ ,\qquad \alpha=-{3\over 2} \ .
	\end{equation}
	Finally, if we define the loop space Dirac delta function
	\begin{equation}
	\delta[C-C_0]
	\equiv \lim_{\epsilon\rightarrow 0}\left({1\over \pi
	\epsilon}\right)^{3/2}\exp\left(-{1\over 2\epsilon}
	\Sigma^{\mu\nu}[C-C_0] \Sigma_{\mu\nu}[C-C_0]\right)
	\end{equation}
	then, the kernel normalization constant is fixed by the boundary
        condition (\ref{bcond}), and we finally obtain the promised
	expression of the quantum kernel as an exact evaluation of the path
	integral
	\begin{equation}
	K[x(s),x_0(s);A]=\left({m^2\over 2i\pi\hbar
	A}\right)^{3/2}\exp\left(
	{im^2\over 4\hbar A} \Sigma^{\mu\nu}[C-C_0]
	\Sigma_{\mu\nu}[C-C_0]\right)\ .
	\label{result}
	\end{equation}
	The above equation, in turn, leads us to the following
	representation of the Nambu--Goto closed string propagator
	\begin{eqnarray}
	&&\int_{x_0(s)}^{x(s)}[Dx^\mu(\sigma)]
	\exp\left\{ -{im^2\over \hbar}\int_\Sigma
	d^2\sigma\sqrt{-{1 \over 2}{\dot x}^{\mu\nu}{\dot x}_{\mu\nu}}\right\}=
	\nonumber\\
	&&\int_0^\infty  dA\, e^{-im^2 A/2\hbar}
	\left({m^2\over 2i\pi\hbar A}\right)^{3/2}
	\exp\left({im^2\over 4\hbar A} \Sigma^{\mu\nu}[C-C_0]
	 \Sigma_{\mu\nu}[C-C_0]\right)\ .\nonumber\\
	\label{prop}
	\end{eqnarray}

	Note that, since no approximation was used to obtain equation
	(\ref{prop}), the above representation can also be interpreted as a
	new {\it definition} of the
	Nambu--Goto path integral. This definition is based on the classical
	Jacobi formulation of string dynamics rather than on the customary
	discretization procedure.

	\subsection{Integrating the path integral}

	As a consistency check on the above result, and in order to
	clarify some further
	properties of the path integral, it may be useful to offer an
        alternative derivation of equation (\ref{prop}) which is
	based entirely on
	the usual gaussian integration technique.
	As we have seen in the previous section, the Feynman amplitude
	can be written as follows
	\begin{eqnarray}
	&&K[x(s),x_0(s);A]=
	\int_{x_0(s)}^{x(s)}[Dx^\mu(\sigma)][Dp_{\mu\nu}(\sigma)]\times
	\nonumber\\
	&&\exp\left\{
	{i\over 2\hbar}\int_{X(\sigma)}p_{\mu\nu}dx^\mu\wedge dx^\nu
	-{i\over 2\hbar}\epsilon_{ab}\int_{\Sigma(\sigma)} d\xi^a\wedge
	d\xi^b H(p)\right\}\ .	\label{jak}
	\end{eqnarray}
	In order to evaluate the functional integral (\ref{jak}), without
	discretization of
	the variables, we enlist the following equalities,
	\begin{eqnarray}
	&&\int_{x_0(s)}^{x(s)}[Dx^\mu(\sigma)]\exp\left\{
	{i\over 2\hbar}\int_{X(\sigma)}p_{\mu\nu}\,dx^\mu\wedge dx^\nu
	\right\}=\nonumber\\
	&&\int_{x_0(s)}^{x(s)}[Dx^\mu(\sigma)]\exp\left\{{i\over 2\hbar}\left[
	\int_{X(\sigma)}d\left(p_{\mu\nu}\,x^\mu dx^\nu\right)-
	\int_{X(\sigma)}x^\mu\,dp_{\mu\nu}\wedge  dx^\nu\right]\right\}
	=\nonumber\\
	&&\delta\left[d\left(p_{\mu\nu}dx^\nu\right)\right]
	\exp\left\{{i\over 2\hbar}
	\int_{\partial X(s)}p_{\mu\nu} x^\mu dx^\nu\right\}\ .
	\end{eqnarray}
	The functional delta function has support on the classical,
	extremal trajectories of the string. Therefore, the momentum
	integration is restricted
	to the classical area--momenta and the residual integration
	variables are the components
        of the area--momentum along the world--sheet boundary
	$p_{\mu\nu}(s)$.
	As a matter of fact, boundary conditions fix the initial and
	final string loops $C_0$ and $C$ but not the conjugate momenta.
	In analogy to
	the point particle case, the classical equations of motion on the
	final world--sheet boundary
	\begin{equation}
	d\left( p_{\mu\nu}dx^\nu\right)\Big\vert_{x=x(s)}=0
	\end{equation}
	require that the three normal components of
	$p_{\mu\nu}$ be constant, i.e. $p_{\mu\nu}x^{\prime\,\nu}(s)=
	{\rm const.}$
	Hence, the functional integral over the boundary momentum
        reduces to a three dimensional, generalized, Gaussian integral
	\begin{equation}
	\int [Dp_{\mu\nu}(\sigma)]\delta\left[d\left(p_{\mu\nu}dx^\nu
	\right)\right]
	(\dots)=\int [dp_{\mu\nu}](\dots)  .
	\end{equation}
	Moreover, the Hamiltonian
	is constant over a classical world--sheet and  can be written
	in terms of the boundary $p_{\mu\nu}$. In such a way,
	the path integral is reduced to the Gaussian integral over the
	three components of $p_{\mu\nu}$ which are normal to the boundary
	\begin{eqnarray}
	&&K[x(s),x_0(s);A]=
	{\cal N}\int [dp_{\mu\nu}] \exp\left\{{i\over \hbar}\left[
	p_{\mu\nu}\int_{\partial X(s)} x^\mu(s) dx^\nu -{A\over 4m^2}
	p_{\mu\nu} p^{\mu\nu}\right]       \right\}=\nonumber\\
	&&{\cal N}\int [dp_{\mu\nu}] \exp\left\{{i\over \hbar}\left[
	{1\over 2} p_{\mu\nu}\Sigma^{\mu\nu}[C-C_0]
	 -{A\over 4m^2} p_{\mu\nu} p^{\mu\nu}\right]\right\} .
	\label{fine}
	\end{eqnarray}
	The integral (\ref{fine}) correctly reproduces the expression
	(\ref{prop}).

	\section{Conclusions}

	The main results of this paper can be summarized as follows.
	Starting with the canonical form of the Schild action for
        a closed, bosonic string, it is possible to formulate the
	Hamilton--Jacobi theory of string dynamics in loop space,
	with the proper area of the string manifold playing the role
	of evolution parameter. The conjugate dynamical quantity is an
        area--Hamiltonian which is quadratic in the corresponding
	momenta so that it is possible to extend to strings, {\it indeed to
	any p-brane}, many of the results which are applicable to a
	relativistic point particle.
	The Feynman path integral for quantum strings is then
	equivalent to the functional wave equation (\ref{waveq}) which
	was derived without
	recourse to a lattice approximation. For the kernel
        (\ref{result}), the path integral, or the corresponding wave
	equation can be solved {\it exactly}. \\
	If, as it is normally done, one starts from a reparametrization
        invariant path integral over the string coordinates, the corresponding
        amplitude describes the propagation of strings with fixed ``energy''
	$E=1/2\pi\alpha'$. The relation between the two amplitudes is
	given by equation (\ref{laplace}). \\
     	Three generalizations of the above results are almost
	straightforward:\\
	i) the wave equation for a closed string coupled to a Kalb--Ramond
	field can be obtained from (\ref{waveq}) through the replacement
	\begin{equation}
	{\delta\over\delta x^\mu(s)}\longrightarrow
	{\delta\over\delta x^\mu(s)}+i\kappa
	B_{\mu\nu}(x)\,x^{\prime\,\nu}(s)\ .
	\end{equation}
	Unfortunately, there is no straightforward way to solve the wave
	equation, or to integrate the path integral, for an arbitrary
        gauge potential; \\
	ii) some more work  is required
            to extend the above formalism to the case of a closed p--brane
	     imbedded into a $D$--dimensional spacetime. However,
	no essential difficulties arise in treating higher dimensional
	extended objects;\\
	iii) once the quantum mechanical propagator
	is known, then one can second quantize the system. If we introduce
	the string wave function $\psi[x(s)]$ as an element of a
	functional
        space of string states, then we can write (\ref{waveq}) as
	\begin{equation}
	i\hbar{\partial\over \partial A} |\psi\rangle =
	H\,|\psi\rangle \ ,
	\end{equation}
         where $H$ is the area--hamiltonian operator.
        Then, the corresponding Green function can be written as follows
	\begin{equation}
	G[x(s),x_0(s);m^2]={i\over\hbar}
	\int_0^\infty dA \langle x(s)|e^{i H A/\hbar}
	| x_0(s)\rangle = \langle x(s)|{1\over H}| x_0(s)\rangle \ ,
	\end{equation}
 	or
	\begin{equation}
	 G[x(s),x_0(s);m^2]={\cal N}\int [D\psi^*][D\psi]
	\psi^*[x(s)]
	\psi[x_0(s)]\exp\left({iS[\psi^*,\psi]/\hbar}\right)
	\ ,
	\end{equation}
	where, $\psi^*[x(s)]$, $\psi[x_0(s)]$ constitute a pair of
	complex functional fields, and
	\begin{equation}
	S[\psi^*,\psi]=
	\int [Dx^\mu(s)]\,\psi^*[x(s)]
	\left[-\hbar^2\left(\int_0^1 ds\sqrt{x^{\,\prime\, 2}}\right)^{-1}
	\int_0^1 {ds\over\sqrt{x^{\,\prime\, 2}}}
	{\delta^2\over \delta x^\mu(s)\ \delta x_\mu(s)} +m^4\right]
	\psi[x(s)]\nonumber\\
	\label{dualr}
	\end{equation}
	is the Marshall--Ramond \cite{mr} action for the dual string model.
	Therefore, by extending the classical Hamilton--Jacobi formulation
	of string dynamics into the quantum domain, one arrives at a
	functional field theory in loop space. \\
	As a final speculative
	remark, it seems worth observing that the functional approach to the
	quantum mechanics of strings is analogous, in several ways, to the
	functional approach to quantum cosmology. For instance, the
	``~wave function of the universe~'' is defined in the functional space
	of all possible 3--geometries, and we suggest that spatial loop
	configurations in string theory play the same role as spatially
	closed 3--geometries in quantum cosmology. Likewise, the
	functional
	Schrodinger equation for strings, at fixed areal--time,
	plays the
	same role as the Wheeler--DeWitt equation in quantum cosmology. If
	this analogy is more than coincidental, then quantum string
	theory in loop space may shed some light on the many dark areas of
	quantum cosmology.

	\end{document}